\documentclass[acmsmall,screen]{acmart}

\usepackage{changes}
\renewcommand{\added}[1]{#1}

\usepackage{multirow}
\usepackage{proof}
\usepackage{xspace}
\usepackage{enumitem}
\usepackage{soul}
\usepackage{fancybox}
\usepackage{arydshln}
\usepackage{float}
\usepackage{changepage}
\usepackage[vlined, linesnumbered]{algorithm2e}
\usepackage{threeparttable}
\usepackage{wrapfig}
\AtBeginDocument{%
  }

\newcounter{listing}

\usepackage{listings}

\setcopyright{cc}
\setcctype{by-nc-nd}
\acmJournal{PACMSE}
\acmYear{2026} \acmVolume{3} \acmNumber{FSE} \acmArticle{FSE189}
\acmMonth{7} \acmDOI{10.1145/3808196}

\lstdefinestyle{cppstyle}{
    language=C++,
    basicstyle=\ttfamily\scriptsize,
    keywordstyle=\color{blue}\bfseries,
    commentstyle=\color{gray},
    stringstyle=\color{brown},
    numbers=left,
    numberstyle=\tiny,
    stepnumber=1,
    numbersep=5pt,
    tabsize=4,
    breaklines=true,
    breakatwhitespace=false,
    showspaces=false,
    showstringspaces=false,
    captionpos=b,
    frame=single,
    framerule=0.5pt,
    rulecolor=\color{black},
    framesep=2pt,                    
    xleftmargin=0pt,                 
    xrightmargin=0pt,                
    framexleftmargin=0pt,            
    framexrightmargin=0pt,           
    aboveskip=1em,
    belowskip=1em,
}

\SetAlCapSkip{1em}
\SetAlCapSty{textnormal}            
\SetAlCapFnt{\small\sffamily}       
\SetAlCapNameFnt{\small\sffamily}   

\newcommand\ans[1]{
\vspace{6pt}
\noindent
\doublebox{
    \begin{minipage}{0.96\textwidth}
      #1
    \end{minipage}
  }
}

\def \tool {\textsc{Ditto}\xspace}

\begin{document}

\title{Compiling Code LLMs into Lightweight Executables}

\author{Jieke Shi}
\email{jiekeshi@smu.edu.sg}
\orcid{0000-0002-0799-5018}
\author{Junda He}
\email{jundahe.2022@smu.edu.sg}
\orcid{0000-0003-3370-8585}
\affiliation{%
  \institution{Singapore Management University}
  \country{Singapore}
}

\author{Zhou Yang}
\orcid{0000-0001-5938-1918}
\affiliation{%
  \institution{University of Alberta \& Alberta Machine Intelligence Institute (Amii)}
  \country{Canada}
}
\email{zy25@ualberta.ca}
\authornote{Corresponding Author, Canada CIFAR AI Chair.}

\author{Chengran Yang}
\orcid{0000-0001-6100-8127}
\affiliation{%
  \institution{Singapore Management University}
  \country{Singapore}
}
\email{cryang@smu.edu.sg}

\author{Mykhailo Klymenko}
\orcid{0000-0002-4641-8977}
\email{Mike.Klymenko@data61.csiro.au}
\author{Thong (James) Hoang}
\orcid{0000-0001-5096-4834}
\email{james.hoang@data61.csiro.au}
\author{Sherry (Xiwei) Xu}
\orcid{0000-0002-2273-1862}
\email{Xiwei.Xu@data61.csiro.au}
\author{Zhenchang Xing}
\orcid{0000-0001-7663-1421}
\email{Zhenchang.Xing@data61.csiro.au}
\affiliation{%
  \institution{CSIRO's Data61}
  \country{Australia}
}

\author{David Lo}
\orcid{0000-0002-4367-7201}
\affiliation{%
  \institution{Singapore Management University}
  \country{Singapore}
}
\email{davidlo@smu.edu.sg}

\renewcommand{\shortauthors}{J. Shi, J. He, Z. Yang, C. Yang, M. Klymenko, T. Hoang, S. Xu, Z. Xing, and D. Lo}

\begin{abstract}
  The demand for better prediction accuracy and higher execution performance in neural networks continues to grow. The emergence and success of Large Language Models (LLMs) have produced many cloud-based tools for software engineering tasks such as code suggestion. Although effective, cloud deployment raises concerns over privacy, latency, and reliance on network connectivity. Running LLMs locally on personal devices such as laptops would address these issues, because it enables offline use and reduces response time. However, local deployment is challenging, since commodity devices lack high-performance accelerators such as GPUs and are constrained by limited memory and compute capacity, which makes it hard to execute large models efficiently.

We present \tool, a framework that optimizes both the model size of Code LLMs and the inference programs that execute them, with a focus on statically-typed languages such as C. Our approach integrates two components. The first is a quantization technique inspired by product quantization, which groups model parameters into per-block codebooks via K-Means clustering and stores each weight as a bit-packed low-bitwidth index. The quantizer further supports a mixed-precision mode that keeps a small number of sensitivity-critical tensors in float32. The second component is a compilation pass integrated into LLVM that automatically detects and replaces unoptimized General Matrix-Vector Multiplication (GEMV) operations, which are the most computationally intensive part of code models, with calls into Basic Linear Algebra Subprograms (BLAS) libraries that are highly optimized for the target hardware. The output of \tool is a compiled executable that runs the selected Code LLM on commodity hardware.

We evaluate \tool on three popular Code LLMs, namely Code Llama, MagicCoder, and OpenCodeInterpreter, \added{achieving up to 10.5$\times$ faster inference, 6.4$\times$ lower memory usage, and 10.5$\times$ lower energy consumption compared with their original inference pipelines}, while preserving accuracy close to the full-precision models, with an average loss of only 0.27\% in pass@1. \tool also outperforms the state-of-the-art \textsc{int8} quantization baseline, achieving up to 6.96\% higher pass@1 accuracy, 2.2$\times$ speedup, and 1.6$\times$ memory reduction, which demonstrates the effectiveness of our approach.

\end{abstract}



\maketitle

\section{Introduction}
\label{sec:intro}

In recent years, Large Language Models (LLMs) have achieved remarkable success across various domains. Among them, specialized LLMs trained on large amounts of source code, commonly known as Code LLMs, have demonstrated exceptional accuracy in automating tasks such as code generation~\cite{jiang2024survey,fan2023large}. These models have become integral to modern software development, since they enable developers to write code more efficiently and effectively~\cite{zhao2024surveylargelanguagemodels,10.1145/3605943}. Their success has led to a variety of tools, such as GitHub Copilot~\cite{githubGitHubCopilot} and Cursor~\cite{cursorCursorCode}, that leverage these capabilities to assist developers in everyday tasks~\cite{zhao2024surveylargelanguagemodels,wang2024software,zheng2023survey}.

While most Code LLM-based applications are deployed on cloud servers to leverage high-performance computing resources, there are notable benefits to running them on personal devices such as laptops, provided that their memory and computational demands can be reduced to levels manageable by commodity hardware. \added{Recent studies show that developers increasingly prefer local execution of Code LLMs over cloud-based alternatives~\cite{11126523,10.1145/3708525,lyu2025myproductivityboosted,SERGEYUK2025107610}, because local execution improves privacy, lowers latency, and supports offline use.} Running Code LLMs locally allows developers to retain control over their code and data, which reduces the risk of exposing sensitive information to third-party services. In addition, local execution removes the dependency on network connectivity, which enables development in environments with limited or no internet access. It also reduces latency by eliminating the data transmission delay to and from cloud servers, a delay that is often affected by network conditions. This matters most for real-time applications such as code completion and debugging, where quick responses are essential. Nonetheless, state-of-the-art Code LLMs typically contain billions of parameters and require 20 to 30 GB of memory to run, which exceeds the capacity of common laptops or desktops with 16 GB of memory, so direct local deployment is impractical without specialized optimization.

To date, only a limited number of studies have investigated how to reduce the memory and computational costs of Code LLMs~\cite{avatar,compressor,10.1145/3611643.3616302,afrin2025quantizationdealbreakerempiricalinsights}. These efforts primarily adopt high-level strategies, such as reducing the number or bit width of model parameters, without addressing the underlying implementation of model inference from a program optimization perspective. Without low-level optimization, such approaches cannot fully support high-performance execution of Code LLMs on personal devices. For example, although an increasing number of inference pipelines are implemented in C or C++, such as llama2.c~\cite{llama2-c}, which was developed by one of OpenAI's founders to run the LLaMA-series models efficiently, their performance often remains suboptimal, because of the lack of systematic low-level optimizations and the frequent reliance on manual tuning for specific hardware. In addition, some implementations depend on specialized high-performance computing hardware, such as GPUs or CPUs with high memory bandwidth, which are not available on most personal devices such as business laptops. As a result, their inference programs still require substantial improvement to achieve efficient execution under constrained hardware resources.

In this paper, instead of treating model optimization solely as a parameter reduction task, we propose a model quantization method accompanied by a compilation-time optimization approach, which frames the compression and acceleration of LLMs as a program optimization problem. We present \tool, a framework for jointly optimizing the model size and the inference program of Code LLMs, with a current focus on C-based implementations such as llama.c~\cite{llama2-c}. As a lightweight and portable pipeline with 18k GitHub stars, llama.c provides efficient LLaMA-series inference on edge devices and has gained wide adoption, so we use it as the deployment baseline that our approach builds upon and further optimizes. During inference, an LLM performs computations involving large numbers of high-precision floating-point parameters. Inspired by the idea of product quantization~\cite{matsui2018survey}, which leverages clustering to reduce the bit width of model parameters, \tool treats the model parameters as a set of vectors, clusters them, and stores only the cluster centroids. The original parameters are replaced with indices that point to their assigned centroids, which reduces the bit width of the parameters substantially. After quantization, \tool automatically synthesizes the inference program for the quantized model.

However, simply reducing the bit width of model parameters is not sufficient for efficient execution on personal devices. The inference of Code LLMs involves numerous General Matrix-Vector Multiplication (GEMV) operations, which are computationally intensive and account for over 80\% of total execution time~\cite{kim2025survey,zeng2025abq}. To optimize this, we design an LLVM compilation pass that automatically detects GEMV operations and replaces them with highly optimized implementations from Basic Linear Algebra Subprograms (BLAS) libraries that are specialized for the target hardware. The combination of quantization and compiler-level optimization yields executables that run Code LLMs efficiently under tight memory budgets while also delivering substantial speedups. We evaluate \tool on a personal device equipped with an Apple M4 CPU, which is commonly found in consumer laptops. The results show that \tool enables LLM inference up to 10.5$\times$ faster than the original inference pipeline on the same device, while reducing memory usage by up to 6.4$\times$ \added{and energy consumption by up to 10.5$\times$}, and preserving accuracy close to that of the original models, with at most a 4.45\% loss in pass@1 and an average loss of only 0.27\%. Furthermore, \tool outperforms an \textsc{int8} quantization baseline, achieving up to 6.96\% higher pass@1 accuracy, up to 2.2$\times$ speedup, 1.6$\times$ lower memory usage\added{, and 2.2$\times$ energy savings}.

\vspace{0.5em}
\noindent{\bf{Contributions.}} Our goal in this work is to close the gaps in Code LLM deployment on personal devices that were discussed above. In summary, we make the following contributions:
\begin{itemize}[leftmargin=*]
    \item We present \tool, a framework that automatically optimizes the execution of Code LLMs, which enables them to run on personal devices with limited memory and computational resources.
    \item We propose a quantization scheme paired with a compiler-inspired rewrite, which reduces the bit width of model parameters and accelerates the inference program while preserving accuracy.
    \item We evaluate \tool on popular Code LLMs and show that it can run these models on personal devices with up to 10.5$\times$ faster inference and 6.4$\times$ lower memory usage compared to the original inference programs, while preserving functional correctness with an average loss of only 0.27\% in pass@1.
\end{itemize}

\section{Preliminaries}
\label{sec:prelim}


\subsection{Code LLMs and Their Inference}
\label{sec:code-llms}

Code LLMs are specialized Large Language Models (LLMs) trained on large amounts of source code~\cite{hou2023large,fan2023large,zhang2023survey,wang2024software}, enabling them to perform a variety of software engineering tasks such as code generation~\cite{xu2022systematic,chen2021evaluating} and summarization~\cite{ahmed2024automatic,li2024machines}, vulnerability detection~\cite{zhou2024large,thapa2022transformer}, and program repair~\cite{jiang2023impact,jin2023inferfix,xia2023automated}. These models are typically based on the Transformer architecture~\cite{vaswani2017attention}, whose main components are self-attention and feed-forward layers. In the feed-forward layers, one of the most computationally intensive operations is the General Matrix-Vector Multiplication (GEMV), which is used to compute their outputs.

\vspace{0.2cm}
\noindent{\bf{General Matrix-Vector Multiplication (GEMV).}}
GEMV is a fundamental linear algebra operation that has regained prominence as a critical building block in neural network inference. In the context of Code LLMs deployed on personal devices, where user queries are typically processed sequentially, GEMV plays a particularly crucial role. During each step of autoregressive generation, the model must multiply large weight matrices by single input vectors. This operation, precisely GEMV, requires streaming entire weight matrices through memory for each generated token, which creates significant memory bandwidth pressure. Without careful optimization, inference latency becomes bottlenecked by memory transfers rather than compute. Formally, GEMV is expressed as:

\begin{wrapfigure}{r}{0.52\textwidth}
\vspace{-15pt}
\hspace{0.1cm}
\begin{minipage}{0.48\textwidth}
\begin{lstlisting}[style=cppstyle, caption={Naïve GEMV implementation.},label=fig:naive_gemv]
void gemv_naive(const float* A, const float* x, float* y, int m, int n,
    float alpha, float beta) {
    for (int i = 0; i < m; ++i) {
        float sum = 0.0f;
        for (int j = 0; j < n; ++j) {
            sum += A[i * n + j] * x[j];
        }
        y[i] = alpha * sum + beta * y[i];
    }
}
\end{lstlisting}
\end{minipage}
\vspace{-15pt}
\end{wrapfigure}

\vspace{0.1cm}
\noindent
\textbf{Definition 2.1.} Let $A$ be a matrix of dimensions $M \times N$ and $x$ be a vector of dimensions $N \times 1$. Let $\alpha$ and $\beta$ be any real values. The general matrix-vector product of $A$ and $x$ produces a vector $y$ of dimensions $M \times 1$ such that:
$y(i) = \beta \cdot y(i) + \alpha \cdot \sum_{k=1}^{N} A(i, k) \cdot x(k)$,
for $i = 1, \dots, M$.

Although a naïve implementation of GEMV is simple, it can not leverage hardware-specific optimizations and is therefore significantly slow. Listing~\ref{fig:naive_gemv} shows a basic row-major implementation of GEMV written in plain C, which serves as a conceptual baseline.

This naïve implementation performs the correct computation but does not benefit from cache blocking, vectorization, or fused multiply-add (FMA) instructions, which are some common optimization techniques~\cite{10.1145/3459010}. It therefore serves as a useful pedagogical reference but is not suitable for high-performance inference workloads.

\begin{wrapfigure}{r}{0.52\textwidth}
    \vspace{-15pt}
\hspace{0.1cm}
\begin{minipage}{0.48\textwidth}
\begin{lstlisting}[style=cppstyle, caption={CBLAS interface for GEMV.}, label=fig:cblas_gemv]
void cblas_dgemv(
const CBLAS_LAYOUT Layout,
const CBLAS_TRANSPOSE trans,
const int m, const int n,
const double alpha,
const double *A, const int lda,
const double *x, const int incx,
const double beta, double *y, const int incy
);
\end{lstlisting}
\end{minipage}
\vspace{-15pt}
\end{wrapfigure}

Many optimization efforts have focused on improving GEMV performance, which has produced a number of highly optimized implementations such as the Basic Linear Algebra Subprograms (BLAS) libraries~\cite{blackford2002updated}. These libraries provide efficient implementations of GEMV and other linear algebra operations by leveraging low-level optimizations and hardware-specific features to achieve high performance, optimizations that are currently not utilized by Code LLM inference programs. Listing~\ref{fig:cblas_gemv} shows the interface for double-precision GEMV in common BLAS libraries, which can be used to accelerate GEMV computations in Code LLMs. The \texttt{Layout} argument specifies whether the result vector $y$ is stored in row-major or column-major order. The \texttt{trans} argument indicates whether the matrix $A$ is transposed, which allows multiplication with vectors in different storage orders without creating copies. The parameters \texttt{m} and \texttt{n} correspond to the matrix dimensions, followed by the scalar factors $\alpha$ and $\beta$ (lines 5 and 8), the base address and leading dimension of $A$ (lines 6 and 7), and the increment values \texttt{incx} and \texttt{incy} for traversing the input and output vectors (lines 7 and 8). In this paper, we focus on optimizing GEMV computations in Code LLM inference programs by replacing the naïve implementation with calls to such optimized libraries, which can significantly improve performance.

\subsection{Model Compression}
\label{sec:model-compression}

Model compression aims to reduce the size of a model, and thereby improve other important metrics such as inference latency, memory usage, and energy consumption. Various techniques have been explored for compressing LLMs, including knowledge distillation~\cite{hinton2015distilling}. As the pioneering work on compressing Code LLMs, Shi et al.~\cite{compressor} propose Compressor, which uses a genetic algorithm to simplify model architectures and applies knowledge distillation to reduce model size to as little as 3 MB. Their compressed models achieve an average 4.23$\times$ improvement in inference latency. Building on this, Shi et al.~\cite{avatar} present Avatar, which searches for Pareto-optimal compressed models that jointly optimize model size, inference latency, energy consumption, and carbon footprint, rather than focusing solely on size. Avatar's compressed models reduce energy consumption by up to 184$\times$, carbon footprint by up to 157$\times$, and inference latency by up to 76$\times$. In a similar line of work, Su and McMillan~\cite{Su2024} apply knowledge distillation to GPT-3.5, producing a smaller and more efficient model for code summarization.

Quantization is another widely used approach to compression, which reduces the bit width of model parameters. This technique has also been applied to Code LLMs. Wei et al.~\cite{10.1145/3611643.3616302} and Kaushal et al.~\cite{kaushal2023lord} adopt quantization and low-rank decomposition, respectively, to improve efficiency in code generation. In particular, they directly quantize model parameters to 4-bit integers by truncating the least significant bits of the original floating-point values. Although simple and effective, this approach treats each weight independently and does not adapt to the distribution of values within each layer. In contrast, \tool adopts a clustering-based quantization strategy that groups model parameters into codebooks and replaces them with low-bitwidth centroid indices, which gives each layer an adaptive representation that follows the actual distribution of its weights.

\section{Methodology}
\label{sec:method}

\subsection{Overview}
\label{sec:overview}

\begin{figure}
    \centering
    \includegraphics[width=1.0\linewidth]{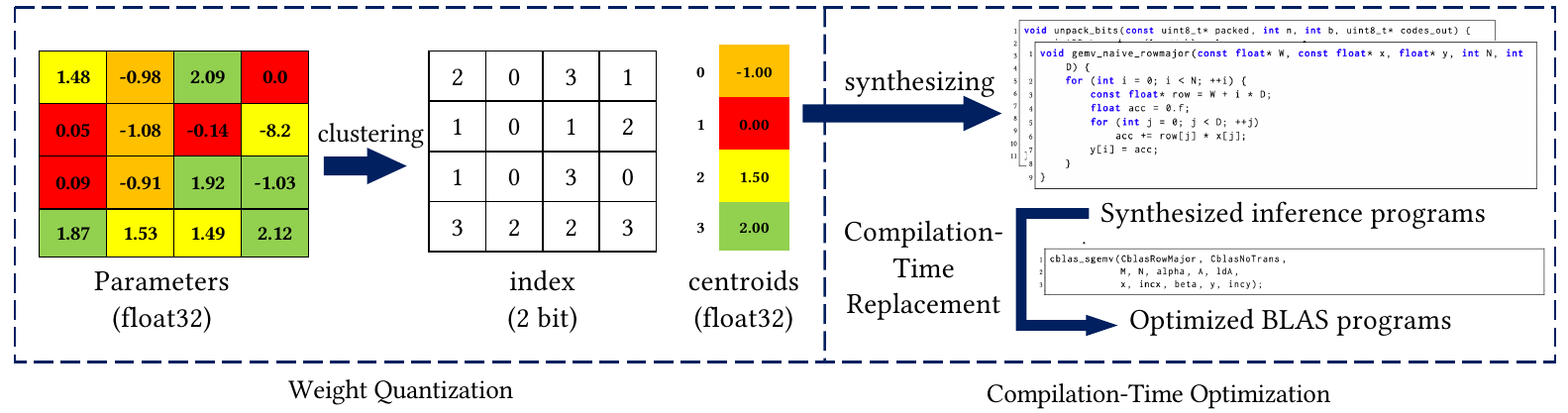}
    \caption{Overview of \tool's two-phase optimization framework.}
    \label{fig:overview}
\end{figure}

\tool is a two-phase framework that combines weight quantization with program-level optimization to accelerate inference for Code LLMs. Figure~\ref{fig:overview} gives an overview of the workflow. The first phase compresses model weights and synthesizes a matching inference program, whereas the second phase optimizes the execution of that program on the target hardware.

In the first phase, \tool applies a K-Means-based quantization scheme that groups weights into a small number of representative values (centroids) and stores each weight as the index of its assigned centroid. The quantizer operates at the sub-block level, which means that each row of a weight matrix is divided into sub-blocks of $G$ consecutive elements, and each sub-block receives its own codebook. This layout adapts the representation to local weight distributions, which is especially beneficial for large models whose attention and projection matrices contain outlier channels. The quantizer also supports a mixed-precision mode that keeps a small number of sensitivity-critical tensors in float32, which lets the user trade a moderate increase in file size for higher fidelity on specific tensors. Centroid indices are serialized with a bit-packing layout that matches the bit width of the indices, so the stored file size is close to the theoretical lower bound. In the second phase, \tool applies a custom LLVM IR pass that identifies GEMV loops in the synthesized inference program and rewrites them into calls to a hardware-tuned BLAS routine such as \texttt{cblas\_sgemv}. The two phases are complementary, since quantization reduces the memory traffic required by each matmul, whereas the compilation pass accelerates the matmul itself.

\subsection{Weight Quantization}
\label{sec:weight-quantization}

\subsubsection{K-Means Clustering with Magnitude Clipping}
\label{sec:clustering}

The first aim of \tool is to represent model weights with a small set of representative float32 values, which amounts to clustering the weights and assigning one representative per cluster. We use 1-D K-Means, because the weights of each layer are scalars and K-Means admits a much faster implementation when its input is one-dimensional. The general K-Means procedure, often referred to as \emph{Lloyd's algorithm}~\cite{lloyd1982least}, alternates between two steps, namely the \emph{assignment} step that maps each data point to its nearest centroid, and the \emph{update} step that moves each centroid to the mean of its assigned points. For $d$-dimensional data with $n$ points and $k$ clusters, the assignment step costs $O(n k d)$ per iteration, because every point is compared against every centroid. In one dimension, however, the optimal clusters are always contiguous intervals in the sorted data. A single sort, done once up front, is therefore sufficient, and each subsequent iteration only needs to update the $k$ interval boundaries rather than revisit every point. This reduces the per-iteration cost to $O(k \log n)$ and makes the algorithm well suited for the hundreds of thousands of small clustering problems that our block-wise quantizer generates.

\tool starts by placing the sorted weights into $k = 2^{b}$ bins of equal population, where $b$ is the target bit width. This is equivalent to a quantile initialization, which naturally allocates more centroids where the weight density is higher. The centroid of each bin is then refined by Lloyd's iteration. Two subtle but important choices drive the quality of the result:

\begin{itemize}[leftmargin=*]
\item \textbf{Efficient centroid update via prefix sums.} After sorting the weights once, we maintain a prefix-sum table so that the mean of any contiguous interval can be computed in constant time. Each iteration locates the new cluster boundaries by binary-searching the midpoints of adjacent centroids within the sorted weight sequence, which keeps the per-iteration cost linear in the number of clusters rather than in the number of weights.
\item \textbf{Magnitude-based scale search.} Let $a = \max_{i} |w_i|$ denote the absolute maximum of the block, which we refer to as the \emph{absmax}. A direct K-Means fit on the raw weights is easily skewed by a handful of outliers near $\pm a$, since placing centroids to cover them diverts budget away from the bulk of the distribution and inflates the overall reconstruction error. To counter this, we introduce a scale factor $f$ that controls how aggressively the tails are suppressed during fitting. For each candidate $f$ in a predefined grid, we clip the weights to $[-f a,\, f a]$ and fit K-Means on the clipped sequence, but perform the assignment step on the \emph{original} unclipped weights, which sends any outlier to the extremal centroid. The value of $f$ that minimizes the reconstruction MSE on the original weights is kept. This decoupling of fitting from assignment frees centroid budget from a few outliers and reallocates it to the bulk of the distribution, which is especially helpful at low bit widths. We use $f \in \{0.5, 0.65, 0.8, 0.9, 0.95, 1.0\}$ as the default grid: it is denser near $1.0$, since the optimum for most rows lies close to the absmax scale, and sparser toward $0.5$ to cover rows with heavier tails. The same six-point grid is used in established low-bit quantizers such as GGML's k-quants~\cite{llamacpp}, and we find that it suffices for every layer we quantize.
\end{itemize}

We also apply an early-exit heuristic. Let $\mu = \tfrac{1}{n}\sum_{i} |w_i|$ denote the mean absolute value of the block. When $a \le 3.5\,\mu$, the block is Gaussian-like, no outliers dominate, and $f = 1.0$ is already optimal. In this case, the entire grid search is skipped. Empirically, this heuristic fires on the majority of blocks in well-behaved layers, which makes the scheme almost free in the common case. Algorithm~\ref{alg:weight-quantization} shows the procedure for a single weight block.

\vspace{0.1cm}
\noindent{\bf{Example.}} Consider a weight block of nine values,
\[
w = [\,0.91,\; 0.92,\; 0.89,\; -0.05,\; -0.06,\; -0.04,\; 1.20,\; 1.21,\; 1.19\,],
\]
and suppose we want to quantize it into three clusters. \tool\ first sorts the values,
\[
s = [-0.06,\; -0.05,\; -0.04,\; 0.89,\; 0.91,\; 0.92,\; 1.19,\; 1.20,\; 1.21],
\]
partitions them into three equal-population clusters, and averages each cluster to obtain the codebook $c = [-0.05,\, 0.91,\, 1.20]$. Each weight is then replaced with the index of its nearest centroid, which yields the sequence $q = [1, 1, 1, 0, 0, 0, 2, 2, 2]$. Since there are only three centroids, each index can be stored in $\lceil \log_2 3 \rceil = 2$ bits. Compared with the original float32 weights, the quantized indices occupy 16$\times$ less memory ($32 \to 2$ bits per element). The codebook itself occupies only $3 \times 32 = 96$ bits and is shared across all weights in the block, so its overhead is negligible whenever the block is large.

\begin{wrapfigure}{r}{0.6\textwidth}
    \vspace{-0.3cm}
\hspace{0.1cm}
\begin{minipage}{0.6\textwidth}
\begin{algorithm}[H]
\small
\SetAlgoLined
\caption{\tool's Weight Quantization Workflow}
\label{alg:weight-quantization}
\DontPrintSemicolon
\KwIn{weights $w \in \mathbb{R}^n$, bit width $b$, grid $\mathcal{F}$}
\KwOut{centroids $c \in \mathbb{R}^{2^b}$, indices $q \in \{0,\ldots,2^b-1\}^n$}
$s \gets \mathrm{sort}(w)$;\; $a \gets \max_i |w_i|$;\; $\mu \gets \tfrac{1}{n}\sum_i |w_i|$\;
\If{$a \le 3.5\,\mu$}{
    $\mathcal{F} \gets \{1.0\}$ \tcp*{Gaussian-like: skip grid}
}
$\mathrm{best\_mse} \gets \infty$\;
\ForEach{$f \in \mathcal{F}$}{
    $s' \gets \mathrm{clip}(s,\, -f a,\, f a)$\;
    Initialize $c$ by quantile of $s'$\;
    \Repeat{convergence}{
        Update cluster boundaries by binary-searching midpoints of $c$ in $s'$\;
        Update centroids via prefix-sum means over $s'$\;
    }
    Assign each $w_i$ to its nearest centroid $c_{q_i}$ and compute MSE on $w$\;
    \lIf{MSE $<$ best\_mse}{record $(c, q, f)$}
}
\KwRet $(c, q)$
\end{algorithm}
\end{minipage}
\vspace{-0.3cm}
\end{wrapfigure}

\subsubsection{Block-wise Granularity and Mixed Precision}
\label{sec:granularity}
Clustering can be applied at several granularities, and the choice directly affects the trade-off between storage overhead and quantization fidelity. The coarsest option is to share a single codebook of $2^{b}$ centroids across an entire row of the weight matrix. This minimizes codebook storage but is fragile: if any column of that row carries outlier values, those outliers consume a disproportionate share of the $2^{b}$ centroids, leaving the remaining weights poorly represented. To contain the damage, \tool instead partitions each row into sub-blocks of $G$ consecutive elements and assigns every sub-block its own codebook. An outlier now only pollutes the centroid budget of its local $G$-element block, so the remaining $(n - G)$ weights in the row are unaffected. The price is a modest increase in storage, namely $(n / G) \cdot 2^{b}$ float32 centroids per row in addition to the bit-packed indices. The user selects $G$ at quantization time, and we find that $G \in \{128, 256\}$ strikes a good balance between codebook overhead and adaptation to local weight distributions. Our quantizer processes all sub-blocks uniformly: the $r \times c$ weight matrix is reshaped into $(r \cdot c / G) \times G$, and the same Lloyd's kernel is invoked independently on each row of the reshaped matrix.

\vspace{0.1cm}
\noindent{\bf{Mixed Precision.}} Some tensors, such as the Q and K projections in the attention block, are highly sensitive to quantization error, because they feed into a softmax whose gradient is large near outliers. In these cases, low-bit quantization can still degrade downstream quality, even at block-wise granularity. To address this, \tool supports a per-tensor-type \emph{fp32 mask}. When the mask flags a tensor type, the quantizer stores that tensor as raw float32 and bypasses clustering entirely. The resulting file uses a mixed layout, where most tensors are quantized and the flagged ones are stored in full precision. This pairs naturally with the compilation-time GEMV pass (Section~\ref{sec:gemv}), since fp32 tensors traverse the naive matmul path and are rewritten into \texttt{cblas\_sgemv} calls by the pass, which means that keeping a tensor in fp32 does not preclude a BLAS-accelerated execution.

\vspace{0.1cm}
\noindent{\bf{Parallel Implementation.}} \tool's quantizer is written in Python and uses Numba's just-in-time (JIT) compiler with \texttt{prange} to parallelize Lloyd's iteration across all sub-blocks of a weight matrix. Each block is sorted once, its prefix-sum representation is reused across all candidates in the scale-factor grid, and the scratch arrays for boundaries and centroids are kept small and thread-local. For a 7B-parameter model, this lets \tool\ complete quantization of all seven weight types across 32 layers in about two minutes on a commodity laptop, without relying on multiprocessing.

\subsubsection{Bit-Packing}
\label{sec:bit-packing}

The output of clustering is a sequence of $b$-bit indices. To realize the memory savings from low-bit representation, these indices must be packed into a dense byte stream, since standard data types are byte-aligned and storing a sub-byte index in an 8-bit container would waste a significant fraction of the storage, for example 50\% for a 4-bit index. \tool\ therefore implements a low-level bit-packer whose format matches the layout expected by the decoder at inference time. Listing~\ref{lst:packbits} shows the packing routine, and Listing~\ref{lst:unpackbits} shows the decoder.

\vspace{0.1cm}
\noindent{\bf{Packing.}} The \texttt{pack\_bits} routine serializes a sequence of $b$-bit indices into a contiguous byte array. A running counter \texttt{bit\_off} tracks the next available bit position in the output. For each index $v$, we need to locate which byte of the output it belongs to and where inside that byte it starts. Since each byte holds $8 = 2^{3}$ bits, the target byte index is $\texttt{bit\_off} / 8$ and the in-byte shift is $\texttt{bit\_off} \bmod 8$. These are computed cheaply as \texttt{j = bit\_off >> 3} and \texttt{sh = bit\_off \& 7}, using the bit-level identities that divide-by-$2^{3}$ equals right-shift by $3$ and mod-$2^{3}$ equals a bitwise AND with $7 = 2^{3}{-}1$. The value $v$ is then shifted left by \texttt{sh} and merged into \texttt{out[j]} using a bitwise OR, which preserves the bits that were already packed. When an index crosses a byte boundary, which is detected by the condition $\texttt{sh} + b > 8$, the overflow bits are shifted right and merged into \texttt{out[j+1]}. The counter \texttt{bit\_off} is then incremented by $b$.

\vspace{0.1cm}
\noindent{\bf{Unpacking.}} \texttt{unpack\_bits} performs the inverse operation. The primary challenge is to read $b$ bits that may begin at any bit offset. The decoder uses a small \emph{sliding window}, formed by combining \texttt{packed[j]} and \texttt{packed[j+1]} into a 16-bit \texttt{chunk}, shifting the chunk right by \texttt{sh}, and masking out the $b$ least-significant bits. This avoids branching on the byte-boundary case and keeps the decoder branch-free. To keep the window safe at the tail of the stream, \tool\ appends a single guard byte to every packed tensor, so the read of \texttt{packed[j+1]} always falls within the allocated buffer.

\begin{figure}[t!]
\centering
\begin{minipage}{0.45\textwidth}
\centering
\begin{lstlisting}[style=cppstyle, caption={Bit-packing $b$-bit codes into bytes.}, label=lst:packbits]
void pack_bits(const uint8_t* codes,
    int n, int b, uint8_t* out) {
  uint32_t mask = (1u << b) - 1u;
  int bit_off = 0;
  for (int i = 0; i < n; ++i) {
    uint32_t v = codes[i] & mask;
    int j  = bit_off >> 3;
    int sh = bit_off & 7;
    out[j] |= uint8_t((v << sh) & 0xFF);
    if (sh + b > 8)
      out[j+1] |= uint8_t(
        (v >> (8 - sh)) & 0xFF);
    bit_off += b;
  }
}
\end{lstlisting}
\end{minipage}
\hspace{1.5em}
\begin{minipage}{0.45\textwidth}
\centering
\begin{lstlisting}[style=cppstyle, caption={Bit-unpacking $b$-bit codes via a 16-bit sliding window.}, label=lst:unpackbits]
void unpack_bits(const uint8_t* packed,
    int n, int b, uint8_t* codes_out) {
  uint32_t mask = (1u << b) - 1u;
  int bit_off = 0;
  for (int i = 0; i < n; ++i) {
    int j  = bit_off >> 3;
    int sh = bit_off & 7;
    uint32_t chunk =
      uint32_t(packed[j]) |
      (uint32_t(packed[j+1]) << 8);
    codes_out[i] = uint8_t(
      (chunk >> sh) & mask);
    bit_off += b;
  }
}
\end{lstlisting}
\end{minipage}
\end{figure}

For bit widths whose packed layout aligns cleanly with byte boundaries, the generic loop is specialized into a fast path that avoids the per-index shift and mask arithmetic. The relevant condition is that a small number of $b$-bit indices fits into an integer number of bytes, which happens whenever $b \cdot k$ is a multiple of $8$ for some small $k$. Three bit widths in particular satisfy this: with $b=8$, each byte is a full index ($k=1$); with $b=4$, two indices fill one byte ($k=2$); and with $b=3$, eight indices fill three bytes exactly ($k=8$, since $3 \cdot 8 = 24$ bits), so the decoder can load three bytes into a 24-bit accumulator and extract all eight indices in a straight-line sequence of shifts. These fast paths are hand-written in \texttt{run\_ditto.c} and together cover every bit width that \tool\ uses in practice, whereas any other bit width falls back to the generic 16-bit sliding window.

\subsection{Inference Kernel}
\label{sec:kernel}

Quantization changes what is on disk; it also changes what the inference kernel must do on every forward pass. Instead of reading a row of float32 weights directly into a dot product, the kernel must first unpack the $b$-bit indices, look them up in a centroid table, and only then compute the dot product with the activation vector. \tool\ ships a hand-tuned quantized GEMV kernel for this purpose, and a small dispatcher that routes each matmul call to either the quantized kernel or to the naive \texttt{matmul()} function that the LLVM pass replaces with \texttt{cblas\_sgemv}.

\vspace{0.1cm}
\noindent{\bf{Quantized GEMV.}} The quantized kernel, shown schematically in Listing~\ref{lst:quantgemv}, iterates over the output rows in parallel using OpenMP. Each iteration unpacks the packed indices of its row into a thread-local float32 scratch buffer, looks the indices up in the corresponding sub-block codebook, and computes a dot product between the reconstructed row and the activation vector.

\begin{wrapfigure}{r}{0.52\textwidth}
\vspace{-0.3cm}
\hspace{0.1cm}
\begin{minipage}{0.50\textwidth}
\begin{lstlisting}[style=cppstyle, caption={Quantized GEMV with per-thread scratch.}, label=lst:quantgemv]
void ditto_matmul(
    float* xout, const float* x,
    const QuantTensor* qt,
    int n, int d, int bits, int k,
    int G, float* scratch,
    int row_stride) {
  #pragma omp parallel for
  for (int i = 0; i < d; ++i) {
    int tid = omp_get_thread_num();
    float* buf = scratch +
        (size_t)tid * row_stride;
    unpack_row(qt, i, n, bits, k, G, buf);
    xout[i] = dot_product(buf, x, n);
  }
}
\end{lstlisting}
\end{minipage}
\vspace{-0.3cm}
\end{wrapfigure}

Two design choices are critical to performance. First, the scratch buffer is allocated once per model as a single contiguous region of size $T \cdot D_{\max} \cdot 4$ bytes, where $T$ is the maximum number of OpenMP threads and $D_{\max}$ is the largest row length that the model uses. Each thread takes a disjoint slab of this region, indexed by \texttt{omp\_get\_thread\_num()}, which eliminates the per-iteration \texttt{malloc}/\texttt{free} that a naive implementation would require. A 7B-parameter model makes roughly 1.4 million GEMV row calls per generated token, so avoiding heap traffic in this loop yields a measurable throughput improvement.

Second, \texttt{unpack\_row} dispatches into a fast path for $b \in \{3, 4, 8\}$ and into a generic 16-bit-sliding-window decoder for all other bit widths. The fast paths exploit the exact byte layout of their bit width, so they avoid shift arithmetic and branch decisions in the hot loop. The inner \texttt{dot\_product} call is hand-vectorized with ARM NEON intrinsics on Apple Silicon and with SSE/AVX on x86, using four independent accumulators to hide the latency of fused multiply-add (FMA) instructions.

\vspace{0.1cm}
\noindent{\bf{Dispatcher.}}
The forward pass does not call \texttt{ditto\_matmul} directly. Instead, it calls a small wrapper, \texttt{tensor\_matmul}, that inspects the \texttt{is\_fp32} flag of the tensor and dispatches either to \texttt{ditto\_matmul} (for quantized tensors) or to the naive \texttt{matmul} function (for fp32 tensors flagged by the mixed-precision mask). The fp32 path is written as a plain doubly-nested loop with no SIMD intrinsics, which is exactly the pattern that the LLVM pass in Section~\ref{sec:gemv} is designed to rewrite. As a result, mixed-precision tensors benefit from both the fidelity of fp32 storage and the speed of BLAS, without any hand-written code in \texttt{run\_ditto.c}.

\subsection{Compilation-Time Optimization}
\label{sec:gemv}

The second phase of \tool is a custom LLVM IR pass that rewrites naive GEMV loops into calls to a tuned BLAS routine. This phase complements the quantization phase, which targets memory traffic, by targeting compute efficiency. The pass is implemented on top of LLVM's \texttt{PatternMatch} API and is loaded into the standard optimization pipeline as a plugin. The pipeline runs in three stages, namely pattern identification, legality and dependency checking, and library replacement.

\vspace{0.2cm}
\noindent{\bf{Stage 1: Pattern Identification.}}
A GEMV pattern in \tool\ is a loop nest of depth at least two, whose innermost body performs a multiply-add reduction over affine memory accesses. Identifying this pattern in LLVM IR is non-trivial, since the same source-level loop can lower into many different IR shapes depending on the optimization level, the target architecture, and the way the induction variable is represented. \tool\ therefore extends LLVM's built-in matchers with three custom combinators. \texttt{OneOf} matches any of a set of alternative sub-patterns, which covers the common variations in affine index computation. \texttt{PHITimesLD} matches the multiplication between a loop induction variable and a leading dimension, which is the canonical form of a row stride. \texttt{AffineFunctionOfPHI} matches more general affine expressions of one or more loop induction variables. These combinators are composed into a higher-level \texttt{GEMVReduction} matcher that looks for the multiply-add reduction itself, and into \texttt{MatchStoreOfVector}, which recognizes the write to the output vector in its many possible forms. Listings~\ref{lst:store_matcher} to~\ref{lst:array_access} show the matcher code.

\begin{minipage}{0.97\textwidth}
\begin{lstlisting}[style=cppstyle, caption={Matcher for storing a GEMV result into the output vector.}, label=lst:store_matcher]
template <typename MatcherType>
auto MatchStoreOfVector(MatcherType GEMVReduction,
                        Value y, Value Alpha, Value Beta,
                        PHINode iv_i, PHINode iv_k,
                        Value LDy, Value GEP) {
    return m_Store(
        OneOf(ScaledVOrV(Alpha, GEMVReduction),
              ScaledVOrV(Beta, m_PHI(m_Value()), GEMVReduction),
              m_c_FAdd(ScaledVOrV(Beta, m_Load(GEP)),
                       ScaledVOrV(Alpha, m_PHI(m_Value()), GEMVReduction))),
        ArrayAccess(y, iv_i, iv_k, LDy));
}
\end{lstlisting}
\end{minipage}

\begin{minipage}{0.45\textwidth}
\begin{lstlisting}[style=cppstyle, caption={Matcher for the multiply-add reduction in the loop body.}, label=lst:gemv_reduction]
inline auto GEMVReduction(
    Value AddLHS, Value MulLHS,
    Value MulRHS, PHINode iv_i,
    PHINode iv_k, Value Alpha,
    Value LDA) {
  return MultiplyAdd(
    Alpha, AddLHS,
    m_Load(ArrayAccess(
      MulLHS, iv_i, iv_k, LDA)),
    m_Load(ArrayAccess(
      MulRHS, iv_k)));
}
\end{lstlisting}
\end{minipage}
\hspace{1.7em}
\begin{minipage}{0.45\textwidth}
\begin{lstlisting}[style=cppstyle, caption={Matcher for array access variants.}, label=lst:array_access]
auto ArrayAccess(
    Value Op, PHINode iv1,
    PHINode iv2, Value LD) {
  return OneOf(
    m_GEP(m_Load(m_GEP(Op, m_PHI(iv1))),
      m_PHI(iv2)),
    m_GEP(m_GEP(Op, PHITimesLD(iv1, LD)),
      m_PHI(iv2)),
    m_GEP(Op, m_PHI(iv1), m_PHI(iv2)),
    m_GEP(Op, AffineFunctionOfPHI(
        iv1, iv2, LD)));
}
\end{lstlisting}
\end{minipage}

\tool's IR pass then scans each function for candidate GEMVs using these matchers, as outlined in Algorithm~\ref{alg:gemv_match}. The pass walks every loop in the function's \texttt{LoopInfo}, selects the ones whose depth is at least two, and applies the matchers to every instruction in the innermost body. When a match succeeds, the pass records the candidate together with the pointers, induction variables, and leading dimension that the matcher has already bound.

\vspace{0.2cm}
\noindent{\bf{Stage 2: Legality and Dependency Checks.}}
Once a candidate is matched, it is subject to a legality check that ensures the rewrite is sound. The core of the check is a data-dependence analysis on the candidate's inner loop, which rejects any loop that contains side effects other than the accumulator update, such as additional stores, aliasing loads, or calls with memory effects. Candidates that fail this check are dropped.

\tool\ also infers the memory layout of the matrix by inspecting the address computation of the matrix load. If the offset is of the form $i \cdot \mathit{ld}_A + k$, the layout is row-major. If it is $k \cdot \mathit{ld}_A + i$, the layout is column-major. Any other pattern causes the candidate to be dropped. In practice, this conservative design produced no compilation failures or silent rewrite errors across all evaluated models, and 100\% of the GEMV operations in the synthesized inference programs matched our canonical pattern and were rewritten successfully.

\begin{wrapfigure}{r}{0.52\textwidth}
\hspace{0.1cm}
\begin{minipage}{0.52\textwidth}
\begin{algorithm}[H]
\small
\caption{LLVM IR pass for locating GEMV candidates.}
\label{alg:gemv_match}
\KwIn{function $F$, loop info $\mathrm{LI}$}
\KwOut{list of GEMV candidates}
\ForEach{loop in $\mathrm{LI}$}{
    \If{loop depth $\geq 2$}{
        Find innermost sub-loops\;
        \ForEach{value $V$ in innermost body}{
            \If{\textup{\texttt{GEMVPattern.Match}}$(V)$}{
                Record candidate\;
            }
        }
    }
}
\end{algorithm}
\end{minipage}
\end{wrapfigure}

\vspace{0.2cm}
\noindent{\bf{Stage 3: Library Replacement.}}
The final stage emits a call to a tuned BLAS routine, such as \texttt{cblas\_sgemv}. The pass inserts the call at the preheader of the candidate's outer loop, using the pointers, dimensions, and leading dimension that were bound in Stage 1. It then removes the store to \texttt{y[i]} in the original inner loop. After the pass finishes, the standard LLVM cleanup passes, namely Dead Code Elimination and CFG simplification, remove the now-unused loop nest. The BLAS symbol itself is declared at IR level using \texttt{Module::getOrInsertFunction}, so \tool's generated IR does not require \texttt{run\_ditto.c} to include \texttt{<cblas.h>}. At link time, the symbol is resolved against the system BLAS, which is Apple Accelerate on macOS and OpenBLAS or Intel MKL on Linux. Listing~\ref{lst:gemv_before} shows a matched GEMV idiom, and Listing~\ref{lst:gemv_after} shows the rewritten code.

\begin{minipage}{0.45\textwidth}
\begin{lstlisting}[style=cppstyle, caption={Matched GEMV idiom in C.}, label=lst:gemv_before]
for (int i = 0; i < M; ++i) {
  float sum = 0.0f;
  for (int j = 0; j < N; ++j)
    sum += A[i][j] * x[j];
  y[i] = sum;
}
\end{lstlisting}
\vspace{-0.3cm}
\end{minipage}
\hspace{1.7em}
\begin{minipage}{0.45\textwidth}
\begin{lstlisting}[style=cppstyle, caption={Rewritten CBLAS call.}, label=lst:gemv_after]
cblas_sgemv(CblasRowMajor,
  CblasNoTrans,
  M, N, alpha, A, ldA,
  x, incx, beta, y, incy);
\end{lstlisting}
\vspace{-0.3cm}
\end{minipage}

\subsection{Implementation}
\label{sec:implementation}

\tool\ is implemented in approximately 4{,}000 lines of Python and C++, including tests and documentation. The quantizer is written in Python with Numba, the inference kernel and the forward pass are written in C on top of llama2.c~\cite{llama2-c}, and the LLVM pass is written in C++ using the new pass manager API. The build pipeline is expressed as a three-stage \texttt{Makefile} target. Stage 1 compiles \texttt{run\_ditto.c} to LLVM IR at \texttt{-O1}, which preserves the loop structure of the naive matmul so that Stage 2 can recognize it. Stage 2 runs \texttt{opt} with the \tool\ pass plugin and rewrites GEMV loops into \texttt{cblas\_sgemv} calls. Stage 3 compiles the rewritten IR to a native executable at \texttt{-O3} with \texttt{-ffast-math} and links against the system BLAS. Model weights and tokenizers are converted from the Hugging Face checkpoint format into llama2.c's native binary layout via an export pipeline, after which \tool\ operates on the converted binaries. We use $b{=}4$ and $G{=}256$ as the default quantization configuration, which empirically provides the best trade-off among accuracy, memory footprint, throughput, and energy consumption.

\section{Evaluation}
\label{sec:evaluation}

\subsection{Experimental Setup}

\vspace{0.1cm}
\noindent{\bf{Platform.}}
We evaluate \tool on a machine with an Apple M4 Chips, 24GB of RAM. We use Clang 14 as the compiler to compile the generated C code into an executable binary. The BLAS library used in our experiments is Apple's Accelerate~\cite{Accelerate}, which is optimized for Apple Silicon chips. The operating system that we use is macOS 13.2.1.

\vspace{0.1cm}
\noindent{\bf{Comparisons.}}
We evaluate \tool by comparing the performance of the optimized programs it generates against two existing implementations as baselines. Note that in the following experiments, we set \tool to use its default 4-bit per-block configuration ($b{=}4$, $G{=}256$) for all models. We leave the exploration of different bit widths to future work.

We first compare \tool against the original floating-point implementation in llama2.c~\cite{llama2-c}, which has been used in previous work~\cite{10638609,10.1145/3611643.3616302}. We chose llama2.c as the baseline for several reasons. First, llama2.c, which was developed by OpenAI co-founder Andrej Karpathy, represents the state of the art in LLM deployment on edge devices and features highly optimized kernel implementations tailored to each hardware platform. Its versatility and robust performance make it an ideal baseline. Second, llama2.c is implemented in plain C and C++ without external dependencies, which ensures maximum compatibility and efficiency across diverse hardware configurations. We use the llama2.c implementation at commit \texttt{d0f8b3c}.

Moreover, we also compare \tool against a strong \textsc{int8} baseline~\cite{10.1145/3611643.3616302}, a widely adopted quantization method that has demonstrated competitive performance in code generation tasks in prior studies~\cite{10.1145/3611643.3616302,afrin2025quantizationdealbreakerempiricalinsights,fang2025smallerweakerbenchmarking}. In our experiments, this baseline is implemented using PyTorch's native post-training dynamic quantization, following the setup of Wei et al.~\cite{10.1145/3611643.3616302} from AWS AI Lab. Concretely, in our \textsc{int8} baseline, all linear layers are quantized from float32 weights into int8 representations, while activations remain in floating-point format. The quantized model is executed with PyTorch's optimized backend, which leverages efficient low-level libraries to accelerate inference on modern CPU architectures. The \textsc{int8} baseline is implemented with PyTorch 2.8.0 with CPU backend.

\vspace{0.1cm}
\noindent{\bf{Models.}}
We evaluate \tool on three popular code-oriented LLMs: Code Llama~\cite{rozière2024codellamaopenfoundation}, Magicoder~\cite{wei2024magicoder}, and OpenCodeInterpreter~\cite{zheng2024opencodeinterpreter}, all of which are open-source and available on Hugging Face~\cite{huggingfaceHuggingFace} and have demonstrated strong performance on code generation tasks across various benchmarks~\cite{evalplus,zhuobigcodebench}.

Code Llama~\cite{rozière2024codellamaopenfoundation} is a code-specialized variant of Llama 2~\cite{touvron2023llama2openfoundation} from Meta, created by further training Llama 2 on code-specific datasets. It can generate both code and natural language descriptions of code, given prompts in either form, and supports multiple programming languages. Magicoder~\cite{wei2024magicoder} is another code-specialized model built on top of Code Llama, further trained on a large corpus of clean and synthetic data, and demonstrates improved performance in code generation. OpenCodeInterpreter~\cite{zheng2024opencodeinterpreter} is a more recent advancement that integrates both code execution and human feedback into its training data, achieving performance comparable to proprietary models such as the GPT-4 series. We use the 7B parameter version of each model, as this size is both widely adopted in practice and represents the upper bound for running full-precision models on a typical laptop with 16 GB or 24 GB of memory. In the following sections, we refer to these models as Code Llama-7B, Magicoder-CL-7B, and OpenCodeInterpreter-CL-7B, following the naming conventions in their official Hugging Face repositories. The 7B versions consist of 32 layers, a hidden size of 4,096, and a feed-forward network dimension of 11,008, with a total size of 28 GB in 32-bit floating point. We use 4-bit per-block quantization for all three models, which is the default setting in \tool.

\vspace{0.1cm}
\noindent{\bf{Benchmarks.}}
We evaluate \tool on two widely used benchmarks for code LLMs: HumanEval+ and MBPP+. HumanEval~\cite{chen2021evaluating} and MBPP~\cite{austin2021program} are among the most commonly used benchmarks for evaluating code generation. Each task in these benchmarks includes a prompt in the form of a task description (e.g., a docstring), and the LLM is expected to generate code that is evaluated against a set of test cases to verify correctness. The original HumanEval benchmark consists of 164 Python programming problems, whereas MBPP includes 974 problems. For a more comprehensive evaluation, we use their extended versions, HumanEval+ and MBPP+, which are enhanced by EvalPlus+~\cite{evalplus} with additional test cases, namely 80 times more for HumanEval and 35 times more for MBPP. We use the official evaluation scripts provided by the benchmarks to assess model performance. The evaluation metrics include pass@1 and pass@5, which measure the percentage of problems for which the generated code passes all test cases on the first and fifth attempts, respectively. We also measure memory usage and inference speed, as these are important non-functional properties for deploying code LLMs on resource-constrained devices.

\vspace{0.1cm}
\noindent{\bf{Research Questions.}}
We aim to answer the following research questions:

\begin{itemize}[leftmargin=*]
    \item \textbf{RQ1:} Can \tool optimize Code LLMs into executables that produce accurate results?
    \item \added{\textbf{RQ2:} How does \tool improve runtime efficiency compared to the original implementation?}
    \item \added{\textbf{RQ3:} What are the individual contributions of weight quantization and compilation-time GEMV optimization to \tool's overall performance?}
\end{itemize}

\subsection{RQ1: Functionality Preservation}

\begin{table*}[t!]
\caption{Performance comparison of full-precision, \tool, and our baseline which runs in \textsc{int8} format for different models on HumanEval+ and MBPP+ datasets.}
\label{tab:quant_results}
\small
\scalebox{0.91}{
\begin{tabular}{cccccc}
\hline\hline
Model & & \multicolumn{2}{c}{HumanEval+} & \multicolumn{2}{c}{MBPP+} \\ \hline
& & Pass@1 & Pass@5 & Pass@1 & Pass@5 \\ \hline
\multirow{2}{*}{Code Llama-7B~\cite{rozière2024codellamaopenfoundation}} & Full-precision & 35.41 & 52.69 & 46.80 & 61.20 \\ \cdashline{2-6}
& \tool & \textbf{36.66 (+1.25)} & \textbf{50.14 (-2.55)} & \textbf{47.29 (+0.49)} & \textbf{56.75 (-4.45)} \\
& \textsc{int8} & 29.70 (-5.71) & 44.57 (-8.12) &  42.41 \added{(-4.39)} & 48.95 (-12.25) \\ \hline
\multirow{2}{*}{Magicoder-CL-7B~\cite{wei2024magicoder}} & Full-precision & 60.45 & 78.55 & 64.20 & 73.29 \\\cdashline{2-6}
& \tool & \textbf{60.68 (+0.23)} & \textbf{76.27 (-2.28)} & \textbf{62.50 (-1.70)} & \textbf{73.54 (+0.25)} \\
& \textsc{int8} & 55.84 (-4.61) & 70.97 (-7.58) & 57.10 (-7.10) & 66.64 (-6.65) \\ \hline
\multirow{2}{*}{OpenCodeInterpreter-CL-7B~\cite{zheng2024opencodeinterpreter}} & Full-precision & 67.70 & 78.44 & 64.20 & 71.84 \\\cdashline{2-6}
& \tool & \textbf{65.87 (-1.83)} & 74.15 (-4.29) & \textbf{62.83 (-1.37)} & \textbf{68.79 (-3.05)} \\
& \textsc{int8} & 65.30 ($-2.40$) & \textbf{76.54 (-1.90)} & 59.88 ($-4.32$) & 67.64 ($-4.20$) \\ \hline\hline
\end{tabular}
}
\end{table*}

\noindent\textbf{Preservation relative to full-precision.}
The results in Table~\ref{tab:quant_results} show that \tool preserves the core code-generation capabilities of state-of-the-art LLMs after 4-bit per-block quantization, with only minor performance degradation in most cases. For pass@1, the preservation is nearly lossless, with an average degradation of just $0.27\%$ across all models, all values within $2\%$, and the largest drop being $1.83\%$ for OpenCodeInterpreter-CL-7B on HumanEval+. In some cases, \tool even improves single-solution accuracy, for example $+1.25\%$ for Code Llama-7B and $+0.23\%$ for Magicoder-CL-7B. These gains align with prior studies~\cite{ganesh-etal-2021-compressing,czachor2007regularization,10.1145/3611643.3616302}, which suggest that quantization can act as a mild form of regularization that reduces overfitting to suboptimal completions. For pass@5, the effect is more noticeable, with reductions up to $4.45\%$ (Code Llama-7B on MBPP+). This suggests that, although the quality of the best solution is largely preserved, quantization narrows the diversity of correct solutions. Nevertheless, the degradation remains low, typically under $3\%$.

\vspace{0.1cm}
\noindent\textbf{Comparison to the \textsc{int8} baseline.}
As Table~\ref{tab:quant_results} shows, \tool also achieves consistent and substantial improvements over the widely used post-training \textsc{int8} quantization baseline. On HumanEval+, \tool delivers up to $+6.96\%$ higher pass@1 and $+5.57\%$ higher pass@5 on Code Llama-7B, whereas on MBPP+, the gains reach up to $+5.40\%$ in pass@1 on Magicoder-CL-7B and $+7.80\%$ in pass@5 on Code Llama-7B. The only exception is pass@5 for OpenCodeInterpreter-CL-7B on HumanEval+, where \textsc{int8} slightly outperforms \tool by $2.39\%$. Overall, \tool outperforms the \textsc{int8} baseline across all models, with average gains of $+4.27\%$ in pass@1 and $+4.06\%$ in pass@5 on each benchmark.

To reduce randomness in code generation, each evaluation is run five times with different random seeds, and we report the average results. We also apply the Mann-Whitney U test~\cite{mann1947test}, a non-parametric statistical test for comparing two independent groups, to measure statistical significance. The results show that \tool significantly outperforms \textsc{int8} in all but one case, with a 99\% confidence level (p-value < 0.01). These findings demonstrate that \tool is more effective than \textsc{int8} at preserving both correctness and solution diversity across all tested models.

\ans{
\textbf{Answer to RQ1:} \tool effectively optimizes state-of-the-art Code LLMs into low-bit programs while preserving their core functionality. The process has a minimal impact on top-1 accuracy (at most a $1.83\%$ drop in pass@1, $4.45\%$ in pass@5), and it consistently outperforms the \textsc{int8} baseline by up to $+6.96\%$ in pass@1 and $+7.80\%$ in pass@5.
}

\begin{table*}[t!]
\centering
\small
\caption{Memory, inference speed, computational efficiency, and energy comparison on Apple M4 CPU.}
\label{tab:performance}
\scalebox{0.96}{
\begin{tabular}{l l c c c}
\hline\hline
Metric & Method & Code Llama-7B & MagicCoder-CL-7B & OpenCodeInterp.-CL-7B \\
\hline
\multirow{3}{*}{Memory (GB)}
 & Full-precision & 26.77 & 26.14 & 26.77 \\
 & \tool          & \textbf{4.2 (6.4$\times$)} & \textbf{4.1 (6.4$\times$)} & \textbf{4.2 (6.4$\times$)} \\
 & int8           & 6.7 (4.0$\times$) & 6.7 (3.9$\times$) & 6.7 (4.0$\times$) \\
\hline
\multirow{3}{*}{Tokens/s}
 & Full-precision & 0.5 & 0.5 & 0.4 \\
 & \tool          & \textbf{3.5 (7.0$\times$)} & \textbf{4.7 (9.4$\times$)} & \textbf{4.2 (10.5$\times$)} \\
 & int8           & 2.2 (4.4$\times$) & 2.1 (4.2$\times$) & 2.2 (5.5$\times$) \\
\hline
\multirow{3}{*}{Latency (ms/tok)}
 & Full-precision & 2000 & 2000 & 2500 \\
 & \tool          & \textbf{286 (7.0$\times$)} & \textbf{213 (9.4$\times$)} & \textbf{238 (10.5$\times$)} \\
 & int8           & 455 (4.4$\times$) & 476 (4.2$\times$) & 455 (5.5$\times$) \\
\hline
\multirow{3}{*}{Eff.\ GFLOP/s}
 & Full-precision & 6.5 & 6.5 & 5.2 \\
 & \tool          & \textbf{45.3 (7.0$\times$)} & \textbf{60.9 (9.4$\times$)} & \textbf{54.4 (10.5$\times$)} \\
 & int8           & 28.5 (4.4$\times$) & 27.2 (4.2$\times$) & 28.5 (5.5$\times$) \\
\hline
\multirow{3}{*}{Energy (J/tok)}
 & Full-precision & 36.0 & 36.0 & 45.0 \\
 & \tool          & \textbf{5.1 (7.1$\times$)} & \textbf{3.8 (9.5$\times$)} & \textbf{4.3 (10.5$\times$)} \\
 & int8           & 8.2 (4.4$\times$) & 8.6 (4.2$\times$) & 8.2 (5.5$\times$) \\
\hline\hline
\end{tabular}
}
\end{table*}

\subsection{\added{RQ2: Memory Usage, Latency, and Efficiency}}

\added{Table~\ref{tab:performance} reports memory footprint, throughput, latency, effective computational efficiency, and energy consumption on Apple M4 CPU. Across all three evaluated models, \tool\ consistently outperforms both full-precision inference and the widely adopted \textsc{int8} baseline.}

\vspace{0.1cm}
\noindent
\textbf{Memory Usage.}
We measure peak memory usage during inference by generating 1{,}024 tokens from a fixed prompt and averaging over five runs.
\tool\ reduces the memory footprint from approximately 26 GB to 4.1 to 4.2 GB for 7B-scale models, which corresponds to a uniform 6.4$\times$ reduction. In comparison, \textsc{int8} reduces memory to around 6.7 GB (3.9 to 4.0$\times$).
The additional savings of up to 1.6$\times$ over \textsc{int8} come from \tool's 4-bit per-block representation. Theoretically, 4-bit quantization provides a compression factor of $2\times$ over int8 and $8\times$ over FP32. In practice, alignment constraints, per-block codebook overhead, and auxiliary data structures such as unpacking buffers moderate these gains, which yields an effective 6.4$\times$ reduction. Even so, this reduction enables 7B-scale models to fit within an 8 GB memory budget, which improves the feasibility of commodity-device deployment.

\vspace{0.1cm}
\noindent
\added{\textbf{Throughput and Latency.}
Using the same experimental setup, we measure throughput in tokens per second (tok/s).
\tool\ achieves 3.5 to 4.7 tok/s across the three models, which represents up to a 10.5$\times$ speedup over full precision.
The \textsc{int8} baseline reaches only 2.1 to 2.2 tok/s (up to 5.5$\times$ speedup).
In terms of latency, this corresponds to 213 to 286 ms per token for \tool, which is up to a 10.5$\times$ reduction compared to full precision (2000 to 2500 ms) and up to a 2.2$\times$ reduction relative to \textsc{int8} (455 to 476 ms).
These improvements stem from two complementary factors, namely (1) reduced memory traffic because of the smaller 4-bit per-block weight footprint and (2) improved kernel efficiency, since our optimized GEMV integrates low-bit unpacking with BLAS-based vectorized computation that amortizes decoding overhead while preserving high parallelism.}

\vspace{0.1cm}
\noindent
\added{\textbf{Computational Efficiency.}}
\added{GFLOPs denote the number of floating-point operations required per token and serve as a hardware-agnostic measure of computational workload. Under the same hardware conditions, a higher GFLOP/s indicates more efficient inference and better hardware utilization. To quantify GFLOPs, we first estimate the per-token cost using a tool provided by Google~\cite{clark2020electra}, which reports that a 7B-parameter transformer requires approximately 12.95 GFLOPs per token during inference. This estimate includes all operations involved in token generation, such as matrix multiplications and activations. Importantly, this cost is invariant to quantization, because regardless of whether weights are stored in full precision or in a compressed format, the number of arithmetic operations in GEMV remains unchanged. Therefore, although quantization reduces memory usage and may improve data-transfer efficiency, it does not alter the fundamental computational workload per token. Based on this constant, we compute effective throughput as $\text{GFLOP/s} = 12.95 \times \text{tok/s}$. \tool\ achieves 45.3 to 60.9 GFLOP/s, which represents up to a 10.5$\times$ improvement over full precision (5.2 to 6.5 GFLOP/s) and up to a 2.2$\times$ improvement over \textsc{int8} (27.2 to 28.5 GFLOP/s), reflecting improved utilization of compute resources.}

\vspace{0.1cm}
\noindent
\added{\textbf{Energy Efficiency.}
We estimate energy consumption per token as $E = P / \text{tok/s}$, where $P$ denotes the sustained system-level power consumption during inference (measured in watts, W). We measure $P$ using the \texttt{powermetrics} utility on macOS, which samples CPU package power at 1-second intervals throughout a 1{,}024-token generation run. Across all configurations, namely full precision, \textsc{int8}, and \tool, the observed power draw remains nearly constant at 18 W, which is expected on the Apple M4. As an integrated SoC, the CPU cores remain fully utilized during inference regardless of weight precision. We therefore use an average power of 18 W for all configurations. Under this model, \tool\ consumes 3.8 to 5.1 J/tok (joules per token), where J denotes energy in joules. This represents up to a 10.5$\times$ reduction relative to full precision (36.0 to 45.0 J/tok) and up to a 2.2$\times$ reduction compared with \textsc{int8} (8.2 to 8.6 J/tok). These results demonstrate substantial improvements in energy efficiency across all evaluated models.}

\ans{
\added{\textbf{Answer to RQ2:}
\tool\ reduces memory usage by up to 6.4$\times$ and accelerates inference by up to 10.5$\times$ relative to full-precision models, while also achieving up to 1.6$\times$ memory savings and 2.2$\times$ throughput improvements over the \textsc{int8} baseline. It further improves effective computational efficiency (up to 60.9 GFLOP/s) and reduces energy consumption by up to 10.5$\times$, demonstrating consistent gains across memory, latency, and energy dimensions.}
}

\subsection{\added{RQ3: Ablation Study}}
\label{sec:ablation}
\begin{table}[t!]
\centering
\small
\caption{Ablation study on Code Llama-7B.
``Quant.\ Only'' applies 4-bit per-block weight quantization without the LLVM-based GEMV optimization.
``BLAS Only'' applies the GEMV-to-BLAS replacement on the full-precision model.
``Full \tool'' combines both phases. Energy is estimated at 18 W sustained power draw.}
\label{tab:ablation}
\scalebox{0.88}{
\begin{tabular}{lccccccc}
\hline\hline
Configuration
& Pass@1
& Pass@5
& Pass@1
& Pass@5
& Memory (GB)
& Tokens/s
& Energy (J/tok) \\
\hline
& \multicolumn{2}{c}{HumanEval+}
& \multicolumn{2}{c}{MBPP+}
& & & \\
\hline
Full-precision
& 35.41 & 52.69 & 46.80 & 61.20 & 26.77          & 0.5            & 36.0 \\
Quant.\ Only
& 36.66 & 50.14 & 47.29 & 56.75 & 4.2 (6.4$\times$) & 0.42 (0.8$\times$) & 42.9 \\
BLAS Only
& 35.41 & 52.69 & 46.80 & 61.20 & 26.77          & 2.4 (4.8$\times$)  & 7.5 \\
Full \tool
& 36.66 & 50.14 & 47.29 & 56.75 & \textbf{4.2 (6.4$\times$)} & \textbf{3.5 (7.0$\times$)} & \textbf{5.1 (7.1$\times$)} \\
\hline\hline
\end{tabular}
}

\end{table}

\added{To isolate the contributions of \tool's two optimization phases, we conduct an ablation study on Code Llama-7B using HumanEval+ and MBPP+. We evaluate three configurations: (1)~\emph{Quantization Only}, which applies 4-bit per-block weight quantization and the synthesized inference program without LLVM-based GEMV-to-BLAS replacement; (2)~\emph{BLAS Only}, which applies compilation-time GEMV optimization to the full-precision model without quantization; and (3)~\emph{Full \tool}, which combines both phases. Table~\ref{tab:ablation} summarizes the results. Because BLAS replacement does not modify the model weights, its accuracy matches that of the full-precision baseline. Similarly, Quantization Only and Full \tool\ use identical quantized weights and therefore achieve the same accuracy.}

\added{The results in Table~\ref{tab:ablation} show that the two phases address distinct bottlenecks and provide complementary benefits. Quantization achieves a 6.4$\times$ memory reduction, compressing the 7B model from 26.77\,GB to 4.2\,GB, which allows it to fit within the capacity of commodity laptops. However, it does not improve throughput, since performance slightly decreases to 0.8$\times$ of the baseline because of the runtime overhead of bit unpacking and centroid lookup during each GEMV computation. In contrast, BLAS replacement yields a 4.8$\times$ speedup by replacing naive loop-based GEMV with hardware-optimized library calls, but it provides no memory savings because the weights remain in full precision. When both phases are combined, \tool\ achieves a 7.0$\times$ speedup, which is substantially higher than the 4.8$\times$ gain from BLAS replacement alone. This improvement arises because quantization reduces the memory bandwidth required for GEMV, which allows the BLAS kernels to operate more efficiently and to enable greater acceleration than either technique alone. The energy column in Table~\ref{tab:ablation} further illustrates this effect. Quantization alone increases energy consumption (42.9 J/tok versus 36.0 J/tok for full precision), since the reduced memory footprint is offset by slower throughput caused by unpacking overhead. BLAS replacement alone reduces energy to 7.5 J/tok (4.8$\times$) by directly improving computational throughput. When the two phases are combined, Full \tool\ achieves 5.1 J/tok (7.1$\times$), which exceeds the gains of either phase individually. These results confirm that quantization and compilation-time optimization are synergistic rather than independent, because quantization alleviates the memory-bandwidth bottleneck, BLAS replacement mitigates the compute bottleneck, and their integration yields amplified improvements in throughput, memory usage, and energy efficiency.}

\ans{
\added{\textbf{Answer to RQ3:}
\tool's two phases address complementary bottlenecks. Together, they achieve a 7.0$\times$ speedup, a 6.4$\times$ memory reduction, and a 7.1$\times$ reduction in energy consumption, outperforming either quantization or BLAS replacement alone due to their synergistic interaction.}
}

\section{Additional Analysis}
\label{sec:additional-analysis}

\subsection{Time Cost of Quantization and Compilation}
\label{sec:compilation-time}

We measure the quantization and compilation time required by \tool\ to generate an optimized executable from a 7B-parameter Code LLM. On our evaluation machine (Apple M4 CPU, 24 GB RAM), the complete pipeline, which includes model quantization, bit-packing, and LLVM-based GEMV optimization, completes in approximately 17 minutes for all three models.
This processing time is relatively short compared with the overall lifecycle of deploying a Code LLM, since it represents a one-time cost per model configuration. Once the optimized binary is generated, it can be reused indefinitely without recompilation, unless the model weights or target hardware change. Moreover, given the sustained performance improvements during inference, which reach up to 10.5$\times$ faster execution and 6.4$\times$ lower memory usage as demonstrated in RQ2, the time overhead is minimal and can be easily amortized over multiple inference runs.

\begin{table}[t!]
\centering
\small
\caption{Comparison of quantization methods on Code Llama-7B (HumanEval+).
AWQ-4bit stores weights in 4-bit format but dequantizes to int8 on CPU
due to the absence of native 4-bit dot-product instructions.}
\label{tab:quant_compare}
\begin{tabular}{lccccc}
\hline\hline
Method & Bits & Pass@1 & Mem.\ (GB) & Tok/s & Energy (J/tok) \\
\hline
Full-precision  & 32 & 35.41                          & 26.8                       & 0.5                       & 36.0 \\
int8            & 8  & 29.70 (-5.71)                   & 6.7 (4.0$\times$)          & 2.2 (4.4$\times$)         & 8.2 \\
AWQ-4bit        & 4  & 28.98 (-6.43)                   & 5.4 (5.0$\times$)          & 2.4 (4.8$\times$)         & 7.5 \\
\tool           & 4  & \textbf{36.66 (+1.25)}          & \textbf{4.2 (6.4$\times$)} & \textbf{3.5 (7.0$\times$)} & \textbf{5.1} \\
\hline\hline
\end{tabular}
\end{table}

\subsection{\added{Comparison with Other Quantization Methods}}

\added{To compare \tool\ with alternative quantization approaches, we evaluate it against Activation-aware Weight Quantization (AWQ)~\cite{AWQ2024}, a state-of-the-art method that maps individual weights to low-bit integers guided by activation statistics and has been shown to preserve accuracy at 4-bit precision for code LLMs~\cite{afrin2025quantizationdealbreakerempiricalinsights}, on Code Llama-7B under the same hardware environment.
AWQ and similar approaches, such as GPTQ~\cite{frantar2023gptq}, require a calibration dataset to reduce quantization error.
In contrast, \tool\ performs clustering-based quantization directly on pretrained weights without calibration or retraining.
Since curating a representative calibration dataset for code LLMs is non-trivial and beyond the scope of this work,
we follow Afrin et al.~\cite{afrin2025quantizationdealbreakerempiricalinsights}
and use the publicly available AWQ 4-bit checkpoint from Hugging Face~\cite{huggingfaceTheBlokeTom}.
Inference is executed on CPU using the official Hugging Face pipeline without additional optimizations.}

\added{Table~\ref{tab:quant_compare} shows that \tool\ outperforms all baselines in accuracy, memory, throughput, and energy. AWQ achieves 2.4 tok/s, which is comparable to int8 (2.2 tok/s), but exhibits a larger pass@1 drop ($-6.43\%$ versus $-5.71\%$). This comes from AWQ's CPU execution path, since in the absence of native 4-bit dot-product instructions, 4-bit weights are dequantized to int8 or fp16 before computation~\cite{huggingfaceGPTQ}. Although storage uses 4 bits, arithmetic proceeds at higher precision, which limits throughput and energy efficiency. Consequently, AWQ consumes 7.5 J/tok, only slightly better than int8 (8.2 J/tok), whereas \tool\ achieves 3.5 tok/s, a $+1.25\%$ pass@1 improvement, and 5.1 J/tok. This advantage arises from tightly coupling 4-bit per-block storage with native bit unpacking and BLAS-optimized GEMV execution, which eliminates intermediate dequantization and reduces both memory traffic and computational overhead.}

\added{More broadly, this comparison highlights a fundamental architectural distinction. GPU-oriented methods such as AWQ~\cite{AWQ2024} and GPTQ~\cite{frantar2023gptq} depend on specialized CUDA kernels for low-bit arithmetic and revert to higher-precision computation on CPUs, which diminishes their theoretical gains in CPU settings. Meanwhile, classical Vector Quantization (VQ)~\cite{rokh2023comprehensive} and its variant Product Quantization (PQ)~\cite{matsui2018survey} share a similar high-level idea with our method, i.e., reducing bit width through clustering. They quantize weight blocks into codebooks and perform inference through lookup-table accumulation. Although memory-efficient, these approaches introduce additional overhead from codebook access and reconstruction. Without optimizing the inference program, this overhead can offset the benefits of compression on CPUs, as shown in our RQ3 ablation study (Section~\ref{sec:ablation}). In contrast, our approach not only compresses weights but also optimizes the inference program. By replacing costly GEMV loops with BLAS-optimized kernels, \tool\ compensates for the overhead of codebook lookups and enables efficient execution of low-bit models on commodity CPUs. This co-design of quantization and program-level optimization provides a technically distinct perspective on efficient LLM inference under tight resource constraints.}

\begin{table}[t!]
\centering
\small
\caption{Generalization of \tool\ across model architectures, scales, and domains.
Speedup and memory reduction are measured relative to full-precision inference.}
\label{tab:generality}
\scalebox{0.90}{
\begin{tabular}{l c c c c c c}
\hline\hline
Model & Arch. & Domain & Params & FP Mem. (GB) & \tool\ Mem. (GB) & Speedup \\
\hline
CodeQwen-1.5-7B~\cite{huggingfaceQwenCodeQwen157BHugging}  & Qwen  & Code    & 7B   & 27.1 & 4.4 (6.1$\times$)  & 3.9$\times$ \\
Code Llama-13B~\cite{rozière2024codellamaopenfoundation}   & LLaMA & Code    & 13B  & 53.0 & 9.7 (5.5$\times$)  & 6.1$\times$ \\
TinyLLaMA~\cite{huggingfaceKarpathytinyllamasHugging}        & LLaMA & General & 110M & 0.45 & 0.11 (4.1$\times$) & 4.2$\times$ \\
\hline\hline
\end{tabular}
}
\end{table}

\subsection{\added{Generalization of \tool}}
\added{A natural question is whether the optimizations in \tool are specific to the LLaMA architecture used in our primary evaluation. To assess generality, we apply \tool to models with different architectures, scales, and domains (Table~\ref{tab:generality}).}

\vspace{0.1cm}
\noindent\added{\textbf{Architecture Generality.}
Both components of \tool, namely weight quantization and GEMV-to-BLAS replacement, operate on standard Transformer linear layers and do not depend on architecture-specific mechanisms. We validate this on CodeQwen-1.5-7B~\cite{huggingfaceQwenCodeQwen157BHugging}, which adopts the Qwen architecture~\cite{bai2023qwentechnicalreport} and differs from LLaMA in its attention mechanism and feed-forward network design. Applying \tool\ to CodeQwen-1.5-7B yields a 6.1$\times$ memory reduction, from 27.1\,GB to 4.4\,GB, and a 3.9$\times$ speedup, which is comparable to the results for Code Llama-7B. These results confirm that \tool's optimizations are not tied to a specific model architecture and can be effectively applied to other Transformer-based LLMs.}

\vspace{0.1cm}
\noindent\added{\textbf{Scale and Domain Generality.}
We further evaluate \tool\ on Code Llama-13B~\cite{rozière2024codellamaopenfoundation} and TinyLLaMA-110M~\cite{huggingfaceKarpathytinyllamasHugging} to examine scalability across model sizes and domains. For Code Llama-13B, memory usage decreases from 53.0\,GB to 9.7\,GB (5.5$\times$) with a 6.1$\times$ speedup. Notably, the full-precision 13B model exceeds the 24\,GB memory capacity of our evaluation machine, whereas the quantized version fits comfortably, which demonstrates that \tool\ enables practical deployment at larger scales. For TinyLLaMA-110M, which is a general-purpose language model rather than a code-specialized one, memory usage decreases from 0.45\,GB to 0.11\,GB (4.1$\times$) with a 4.2$\times$ speedup. This result indicates that \tool's optimizations are not limited to code-specific models and generalize to other domains as well. Taken together, these findings confirm that \tool's two-phase approach is broadly applicable across model scales and application domains.}

\vspace{0.1cm}
\noindent\added{\textbf{Hardware Portability.}
Although our evaluation is conducted on an Apple M4 CPU, \tool\ is not platform-specific. The quantization operates at the model level and produces a platform-independent compressed model. The GEMV-to-BLAS replacement is implemented as an LLVM IR pass that emits calls to the standard CBLAS interface. At link time, these calls bind to the local BLAS provider, such as Apple Accelerate, OpenBLAS, or Intel MKL. Since optimized BLAS libraries are available for modern CPU architectures including x86, ARM, and RISC-V, the same compiled pipeline can be deployed by relinking against the target platform's BLAS implementation. Memory reduction remains consistent across platforms, whereas throughput improvement depends on the quality of the underlying BLAS library.}

\subsection{\added{Implications for Software Engineering}}

\noindent
\added{\textbf{Takeaway 1 (for software developers):} With \tool, developers can run Code LLMs locally on standard laptops without GPUs, achieving latency below 300\,ms with approximately 4\,GB of memory. This enables fast, privacy-preserving code completion and analysis directly within IDEs, without transmitting proprietary code to cloud services.}

\vspace{0.1cm}
\noindent
\added{\textbf{Takeaway 2 (for engineering teams):}
Teams can integrate \tool\ as a build-time step that converts pretrained models into compact, GPU-free C executables in approximately 17 minutes (Section~\ref{sec:compilation-time}). The resulting binary depends only on a system BLAS library, which facilitates seamless integration into CI/CD pipelines and supports on-device deployment of AI coding assistants.}

\vspace{0.1cm}
\noindent
\added{\textbf{Takeaway 3 (for SE researchers):}
\tool\ illustrates that co-designing model compression with program-level optimization yields synergistic gains that exceed those of either technique alone (Section~\ref{sec:ablation}). This principle can guide future research on efficient LLM-powered software engineering tools, such as on-device program repair and privacy-sensitive code search.}

\subsection{Threats to Validity}
A potential threat to internal validity arises from randomness in code generation and quantization. To mitigate this issue, we repeated each experiment five times with different random seeds and reported the average results, following Arcuri and Briand~\cite{10.1145/1985793.1985795}. Although clustering in quantization depends on centroid initialization, our iterative refinement process converges to a stable configuration, which reduces sensitivity to randomness. Another concern is the correctness of our custom LLVM pass, which we addressed through extensive unit testing and by verifying functional equivalence between the original and optimized code on representative examples. Regarding external validity, we evaluated three widely used code-specialized LLMs on two standard benchmarks using an Apple M4 CPU. These models, datasets, and hardware are representative of common usage scenarios for Code LLMs. Nevertheless, future work could further examine generalizability across additional models, benchmarks, and platforms. For construct validity, our metrics, including pass@1, pass@5, inference speed, and memory usage, are standard in LLM evaluation and directly reflect practical usability.

\section{Related Work}
\label{sec:rel-work}

\added{The first line of related work focuses on model compression~\cite{10.1145/3708525,lo2023trustworthy}, which reduces the size of LLMs to improve inference latency, memory usage, and energy consumption. As an early effort on compressing Code LLMs, Shi et al.~\cite{compressor} use a genetic algorithm to simplify model architectures and apply knowledge distillation to shrink models, which achieves an average $4.23\times$ latency improvement. They subsequently introduce Avatar~\cite{avatar}, which jointly optimizes model size, latency, energy consumption, and carbon footprint. Similarly, Su and McMillan~\cite{Su2024} distill GPT-3.5 into a smaller model for code summarization. Other approaches include quantization~\cite{10.1145/3611643.3616302} and low-rank decomposition~\cite{kaushal2023lord}, both of which achieve notable efficiency gains for code generation. Recent work further examines the non-functional properties of compressed Code LLMs, including robustness~\cite{fang2025smallerweakerbenchmarking}, privacy~\cite{haque2025quantizationimpactsprivacyrisk}, and output code quality~\cite{afrin2025quantizationdealbreakerempiricalinsights}.}

\added{There are also efforts on software-hardware co-design for efficient LLM inference~\cite{xu2024vcllmvideocodecssecretly,mo2025lut,park2024lut,wei2025t}, which jointly optimize the model and the underlying hardware architecture to improve performance. Some studies target high-performance platforms such as GPUs or ASIC accelerators, including VQ-LLM~\cite{huang2025vqt}, MVQ~\cite{li2025mvq}, and VQT-CiM~\cite{huang2025vqt1}. VQ-LLM depends on custom CUDA kernels, MVQ relies on GPU tensor cores, and VQT-CiM targets ferroelectric compute-in-memory accelerators. These hardware-specific implementations differ from our focus on commodity CPU-only devices without such hardware support. In addition, neural codebook methods such as QINCo2~\cite{vallaeys2025qinco2vectorcompressionsearch} address a fundamentally different problem, which is compressing input vectors into quantized embeddings for similarity search rather than compressing model weights for LLM inference. Such methods also use neural networks to construct codebooks and reconstruct vectors, whereas \tool\ performs table lookup to recover weights and co-designs quantization with compilation-time GEMV-to-BLAS replacement. This integration enables low-bitwidth execution while also systematically optimizing the inference program itself, which has not been explored in prior work. We therefore believe that our method is both technically novel and provides a unique perspective in this field.}

Program-level optimization of Code LLMs has also been explored, primarily by restructuring inputs or reducing token counts to lower processing costs, since inference time scales with input length. Token pruning methods remove less important tokens to shorten sequences~\cite{cigar,Zhang2022diet,10.1145/3650212.3680347}. For example, Zhang et al.~\cite{Zhang2022diet} propose DietCode, which selects high-attention statements and tokens to reduce the computational cost of Code LLMs. Although effective at shrinking inputs, these approaches do not optimize the underlying inference program itself, which limits overall system-level gains. In contrast, our work directly improves the low-level execution of the inference pipeline and complements input-level reductions with compiler-based acceleration.

Beyond efforts specific to Code LLMs, compiler research has long leveraged pattern matching and LLVM IR passes for program optimization, such as LoopTactics~\cite{10.1145/3372266}, KernelFaRer~\cite{10.1145/3459010}, and MMTR~\cite{10.1145/3719276.3725197}, which perform nested loop and matrix multiplication optimizations through LLVM-based idiom recognition and rewriting. However, these systems target general-purpose program optimization rather than LLM inference workloads. Our work instead introduces a specialized framework tailored to Code LLM inference, which provides a focused and high-performance solution for this emerging domain.

\section{Conclusion and Future Work}
\label{sec:conclusion}

\added{In this work, we present \tool, a synthesis- and compiler-inspired framework for optimizing Code LLM inference programs. By combining K-Means-based weight quantization with an LLVM-based GEMV rewrite, \tool\ generates low-bitwidth, memory- and compute-efficient executables that run on commodity CPUs. When evaluated on Code Llama, MagicCoder, and OpenCodeInterpreter, \tool\ achieves up to a 10.5$\times$ speedup and a 6.4$\times$ memory reduction over the original pipelines while maintaining near full-precision accuracy, with an average pass@1 loss of only 0.27\%. Compared with an \textsc{int8} baseline, it delivers up to 6.96\% higher pass@1, a 2.2$\times$ speedup, and 1.6$\times$ lower memory usage. The end-to-end quantization and compilation process takes only 17 minutes for a 7B model, which makes deployment practical on personal devices. Currently, \tool\ targets inference pipelines based on the llama.c design, because of its popularity and lightweight, dependency-free structure, which enables fine-grained control over model loading and kernel replacement. In future work, we plan to extend support to other implementations, such as llama.cpp and PyTorch or TensorFlow pipelines, and to broaden the optimization pass to additional kernels so that it can enable batch processing and support encoder-decoder and multimodal architectures.}

\section*{Data Availability}

The code of \tool\ is available at \url{https://github.com/jiekeshi/ditto}.

\section*{Acknowledgments}
This research / project is supported by the A*STAR under its 2nd CSIRO and A*STAR: Research-Industry (2+2) Partnership Program (Award R24I5IR047). Any opinions, findings and conclusions or recommendations expressed in this material are those of the author(s) and do not reflect the views of the A*STAR. We would also like to thank the anonymous reviewers for their valuable feedback and suggestions.

\bibliographystyle{ACM-Reference-Format}
\bibliography{ref}

\end{document}